\documentclass{article}
\usepackage{spconf,amsmath,graphicx,hyperref}
\usepackage{url}
\usepackage{xurl}
\usepackage{verbatim}
\usepackage{color}

\title{Coordinated Electromagnetic Side-Channel Attacks for Voter--Ballot Linking: A Case Study of the Brazilian E-Polling System}
\name{Leandro Hyeda$^{1}$,  Kleber V. Cardoso$^{2}$,
Antonio Oliveira-Jr$^{2,3}$, Saulo Queiroz$^{1}$\thanks{This
work has been partially funded by the project Advanced Multimodal Sensing (AIMS),
supported by the Advanced Knowledge Center in Immersive Technologies (AKCIT),
with financial resources from the PPI IoT of the MCTI, grant number 057/2023,
signed with EMBRAPII, and by the Fundação de Amparo à Pesquisa do Estado
de Goiás (FAPEG), research grant 64448878/2024.
}}

\address{
$^{1}$Federal University of Technology -- Paraná (UTFPR),
Ponta Grossa -- PR, Brazil\\
$^{2}$Federal University of Goiás (UFG),
Goiânia -- GO, Brazil\\
$^{3}$Fraunhofer Portugal AICOS, Porto 4200-135, Portugal\\
E-mail: leandrohyeda@alunos.utfpr.edu.br;
\{kleber, antoniojr\}@ufg.br;\\
sauloqueiroz@utfpr.edu.br
}

\begin{document}

\onecolumn
\maketitle

\begin{abstract}
In this work, we show how two adversaries (\emph{Eve} and \emph{Mallory})
can coordinate an electromagnetic side-channel attack to reconstruct the
vote displayed on an e-voting machine (EVM) and link it to a specific voter
(\emph{Alice}). Assuming that Eve has access to a place adjacent to the 
e-polling room (e.g., restroom, unsupervised room), she performs vote reconstruction 
by intercepting unintended electromagnetic emanations associated with the video signal.
Meanwhile, \emph{Mallory} observes when \emph{Alice} is voting
and reports this information to \emph{Eve} over the Internet. To validate the most critical 
stage of the attack, we conduct a software-defined radio experiment using the official Brazilian
 e-voting interface and demonstrate through-wall vote reconstruction. Although conventional 
monitors are employed rather than official EVMs, our findings
provide insights into information forensics and support security recommendations
for government authorities responsible for e-voting systems.
\end{abstract}

\begin{keywords}
Side-channel attack, TEMPEST, Electromagnetic Leakage,  e-voting,
Brazilian electoral system.
\end{keywords}

\section{Introduction}

Ballot secrecy is a fundamental requirement for free elections in democratic countries.
In highly populated countries such as Brazil, electronic voting (e-voting)
systems can significantly accelerate the reporting of election results.
However, e-voting systems are subject to specific classes of threats that may
compromise ballot secrecy. In this context, electromagnetic side-channel
attacks against displays---often referred to as TEMPEST attacks---represent a relevant threat, as they may allow an adversary to reconstruct sensitive
screen information by remotely intercepting unintended electromagnetic
emanations resulting from the transmission of video signals within the 
target system~\cite{tempestclass-ieeeaccess2025},~\cite{lvds-ieeetifs-25},~\cite{vaneck1985tempest}.

Electromagnetic threats to ballot secrecy have been investigated in earlier studies. 
In~\cite{gonggrijp2006nedap}, the authors experimentally analyzed compromising emanations 
from the Nedap-Groenendaal ES3B voting machine, demonstrating that received radio signals could 
reveal information about electoral choices, although no reconstruction of the displayed e-vote image is reported.
Their experiments included distinguishing choices through 
display refresh behavior and examining candidate-dependent display data bursts. 
The Dutch TEMPEST controversy and its implications for electronic voting are further 
discussed in~\cite{dutchtempest-2009}. Similarly, the authors of~\cite{evote-components-2011} discuss how 
common EVM components may be vulnerable to TEMPEST attacks, but do not report reproducible experiments.

More recently, the authors of~\cite{publicTEMPEST2026} introduced the concept of ``public TEMPEST'' 
to describe TEMPEST attacks facilitated by publicly available information about a public system.
They use the publicly available official 
interface of the Brazilian e-voting system to characterize its spectral pattern before acquiring 
the corresponding electromagnetic leakage. Building on this concept, the authors of~\cite{wticg2026} 
show how public information about an e-voting system can reduce the computational complexity of frame 
refresh rate estimation from $O(N\log N)$ to $O(N)$ for an $N$-sample electromagnetic leakage signal.

In~\cite{sbseg2026}, the authors describe how an adversary could bring a receiver within 0.5~m of the
 EVM to carry out a TEMPEST attack, despite the electromagnetic protection requirements associated with 
the Brazilian UE2020 and UE2022 models~\cite{ue2020security}.
Drawing on a real case in which a city council candidate recruited voters to covertly record their own votes 
using a microcamera embedded in smart glasses, the authors describe a scenario in which voters are recruited 
to carry portable, battery-operated software-defined radios, such as the Ettus USRP E312, 
into the voting booth to record the electromagnetic leakage for subsequent offline analysis.
These prior works do not investigate how a TEMPEST attack could compromise ballot secrecy by linking a reconstructed 
vote to a specific, non-colluding voter (\emph{Alice}).

To fill this gap, we investigate a coordinated attack in which one adversary (\emph{Eve}) 
reconstructs the displayed e-vote from unintended electromagnetic emanations associated with the EVM's 
video signal, while another adversary (\emph{Mallory}) observes when Alice votes and relays this timing 
information to \emph{Eve}, enabling her to link the reconstructed vote to \emph{Alice}.

The remainder of this work is organized as follows. Section~\ref{sec:background} provides 
background on the TEMPEST signal. Section~\ref{sec:attack} describes the coordinated attack,
 and Section~\ref{sec:case} evaluates its viability based on software-defined radio experiments. 
Finally, the last section concludes the work and outlines recommendations and directions for future research.

\begin{figure}[tb]
    \caption{Illustration of a Brazilian polling station (Source: Paulo Pinto/Agencia Brasil~\cite{camara2025polling}).
Although the EVM screen is concealed, the voter using the booth may remain visible to observers near the entrance, potentially facilitating voter--ballot linking.
}   \centering
    \label{fig:epolling}
    \includegraphics[width=7cm]{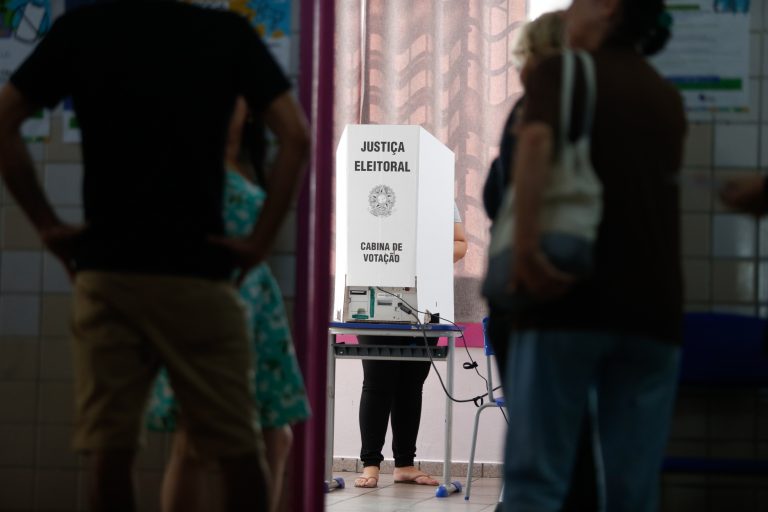}
 \end{figure}

\section{The TEMPEST Signal}\label{sec:background}
In TEMPEST attacks, an eavesdropper reconstructs sensitive information from electromagnetic
 signals unintentionally emitted by the target device~\cite{tempestclass-ieeeaccess2025},~\cite{lvds-ieeetifs-25}. 
In VGA, the signal on a color channel of the video cable can be modeled as rectangular pulses $p(t-nT_p)$ weighted by 
pixel intensities $x[n]$\footnote{For HDMI, a similar pulse model applies to the encoded serial symbols 
rather than directly to pixel intensities~\cite{grtempest-2022}.}
\begin{equation}
x(t) = \sum_{n} x[n]p(t - nT_p), \label{eqn:signal}
\end{equation}
where the pulse (pixel) duration $T_p$ relates to the
pixel rate $f_p$ as follows
\begin{eqnarray}
T_p &=& \frac{1}{f_p} \quad \textrm{s}, \label{eqn:tp} \\
f_p &=& P_x P_y f_v \quad \textrm{pixels/s} \label{eqn:pixelrate},
\end{eqnarray}
and $P_x$, $P_y$, and $f_v$ 
represent the number of pixels per line (including blanking pixels), 
the number of lines per frame (including blanking lines), and the frame rate, 
respectively. The Fourier transform of this idealized waveform $x(t)$ is given by
\begin{equation}
X(f)=P(f)\sum_{n} x[n]e^{-j2\pi nfT_p},
\end{equation}
where $P(f)$ denotes the spectrum of the pulse-shaping function $p(t)$, and
$\sum_{n} x[n]e^{-j2\pi nfT_p}$
is the discrete-time Fourier transform (DTFT) of the pixel sequence $x[n]$. 
The DTFT term is periodic in $f$ with period $f_p$, while $P(f)$ weights the resulting spectral replicas.
For ideal rectangular pulses of duration $T_p$, the spectral envelope vanishes at $f=kf_p$ for every nonzero 
integer $k$, but can remain nonzero at nearby frequencies.
 These isolated nulls do not imply an absence of energy in the surrounding frequency bands.
 In practice, the radiated spectrum also depends on the display circuitry, coupling mechanisms, 
and deviations from the ideal pulse shape. Frequency bands centered at pixel-clock harmonics 
can therefore be investigated as candidates for signal interception.

When the pixel rate can be reasonably estimated a priori, the harmonic frequency
search space of a TEMPEST attack can be substantially reduced. The adversary can
then focus on identifying a suitable compromising harmonic and processing the
corresponding electromagnetic leakage, followed by signal resynchronization using
known display timing parameters to reconstruct the video frames. 

\begin{figure}[tb]
\caption{Coordinated TEMPEST attack for voter--ballot matching.}
    \label{fig:coordinatedtempest}
\centering
    \includegraphics[width=7cm]{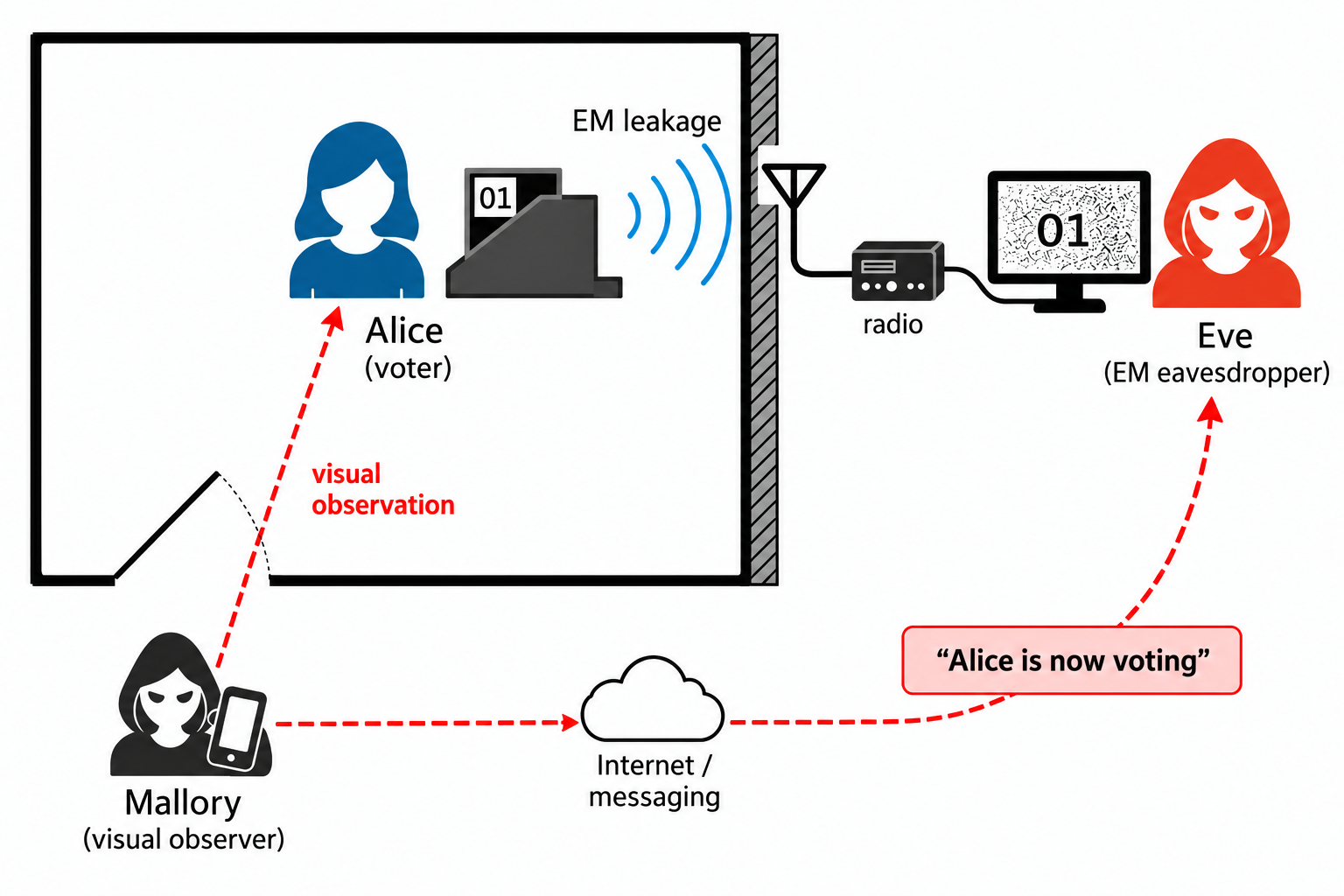}
 \end{figure}

\section{The Coordinated E-Polling TEMPEST Attack}\label{sec:attack} 
In this section, we show how two adversaries can coordinate a TEMPEST
attack against an e-polling section to reconstruct an e-vote image and
associate it with a specific voter. To this end, we first consider the
operational rules of the Brazilian e-polling system, as described in
Section~\ref{subsec:braziliansystem}. Then, in
Section~\ref{subsec:coordinatedtempest}, we describe how the adversaries
can coordinate their actions to perform voter--ballot matching, i.e.,
to associate a reconstructed vote with the voter who cast it.

\subsection{The Brazilian Electronic Voting System} \label{subsec:braziliansystem}
Brazil holds elections every two years, alternating between municipal and national
elections. Voting takes place on a designated election day, with polling stations
open from 8 a.m. to 5 p.m. for voters to cast their ballots in person
at their assigned electoral sections.  Electoral sections are
typically established in rooms of public facilities, such as schools and
universities, designated by the Brazilian electoral authorities for this purpose.

Each electoral section is operated by designated poll workers and may also be
monitored by accredited party poll watchers. Voters typically wait in a queue
outside the electoral section and enter the polling room when authorized. Once
inside, the voter first undergoes an identification procedure conducted by the
poll workers. After successful identification and authorization, the voter
proceeds individually to the voting booth to cast their ballot using the EVM.
To preserve ballot secrecy, the EVM is placed inside a voting booth and positioned 
so that its screen cannot be observed by other individuals in the polling room, as 
illustrated in Fig.~\ref{fig:epolling}. However, the voter using the booth may remain 
visible to observers near the entrance. Two adversaries could exploit this visibility to 
link a reconstructed vote to a specific voter through a coordinated TEMPEST attack, as discussed 
in Section~\ref{subsec:coordinatedtempest}.

\subsection{Coordinated TEMPEST Attack}\label{subsec:coordinatedtempest}
We identify a coordinated attack in which two adversaries, referred to as
\emph{Eve} and \emph{Mallory}, cooperate to associate the reconstructed
e-vote with the voter who cast it. We refer to the target voter as
\emph{Alice}. Fig.~\ref{fig:coordinatedtempest} illustrates the proposed coordinated
TEMPEST attack for voter--ballot linking.
In this attack, \emph{Eve} is responsible for eavesdropping
on the electromagnetic emanations of the EVM and reconstructing
\emph{Alice}'s vote. To this end, \emph{Eve} accesses an unsupervised
location adjacent to the room hosting the electoral section, such as a
restroom, public sidewalk, or neighboring room. From such a location,
however, \emph{Eve} may be unable to determine who is currently casting
a vote at the EVM.

To overcome this limitation, \emph{Mallory} seeks a position from which
the flow of voters entering and leaving the polling room can be observed.
For instance, \emph{Mallory} may be a voter registered in the same
electoral section and wait in the voting queue until being admitted to
cast a ballot. Alternatively, \emph{Mallory} may pose as someone
accompanying a voter and remain in an area from which the entrance to
the polling room is visible. 
Therefore, depending on the layout of the polling location, \emph{Mallory} may 
obtain a line of sight to voters entering the polling room without having access
to the EVM itself. This possibility is also illustrated in Fig.~\ref{fig:epolling}, 
where the polling room layout may allow people near the entrance to identify the voter
 using the voting booth. Poll workers may overlook this exposure if they are unaware that 
observing voter activity could facilitate a coordinated TEMPEST attack, even when the EVM 
screen remains concealed.

From this position, \emph{Mallory} observes when the target voter
\emph{Alice} proceeds to the voting booth and communicates this information
to \emph{Eve} over the Internet. \emph{Eve} can then temporally
correlate \emph{Mallory}'s observation with the electromagnetic leakage
being acquired from the adjacent location. This coordination enables
the adversaries to associate the reconstructed e-vote with
\emph{Alice}, thereby achieving voter--ballot matching.

\section{Attack Viability}\label{sec:case}

Electromagnetic eavesdropping of the e-vote constitutes the most technically
challenging stage of the coordinated TEMPEST attack described in
Section~\ref{sec:attack}. In this section, we experimentally investigate the
viability of this stage from the perspective of \emph{Eve}. Specifically, we
evaluate whether \emph{Eve} can acquire electromagnetic emanations from a
display and exploit them to reconstruct the e-vote image under conditions
representative of the attack scenario in a reasonable time interval.

\subsection{Identification of Compromising Frequency}

To perform an electromagnetic side-channel attack, \emph{Eve} must estimate
the pixel rate of the target display in~(\ref{eqn:pixelrate}) to identify
candidate frequencies for radio acquisition. In a conventional attack, this
typically requires spectral analysis over a range of potential frequencies.
For example, consider a legacy $800\times 600$ display operating at a nominal rate of 60~Hz.
When the blanking intervals are included, the corresponding total timing
dimensions are $P_x=1056$ and $P_y=628$, according to the VESA
standard~\cite{vesa_dmt_2013}. The resulting pixel rate is therefore
$1056\times 628\times 60\approx 39.79$~MHz, providing a reference frequency for 
identifying potentially compromising spectral components. Nevertheless, identifying 
such components within a wide frequency search space can be a challenging stage of a 
TEMPEST attack.

However, in public systems such as the Brazilian e-polling system, information
about the EVM display may already be available in public hardware
specifications issued for procurement or technical testing
purposes~\cite{publicTEMPEST2026}, \cite{sbseg2026}. Such prior knowledge reduces the
number of unknown parameters that \emph{Eve} must infer from the
electromagnetic leakage and, consequently, narrows the frequency search space.
Rather than blindly searching a broad spectrum, \emph{Eve} can prioritize
candidate frequencies and harmonics derived from publicly known display
parameters. For operational Brazilian EVM models, publicly documented
resolutions include $1280\times768$ for the UE2013 and UE2015
models~\cite{ue2015hardware}, and a minimum resolution of $1280\times 720$
for the UE2020 and UE2022 models~\cite{ue2020hardware}.

\subsection{Experimental Validation}
\begin{figure}[tb]
    \caption{E-vote ``91'' reconstructed by \emph{Eve} from through-wall, real-time electromagnetic leakage capture.}
    \label{fig:91}
 \centering
    \includegraphics[width=7cm]{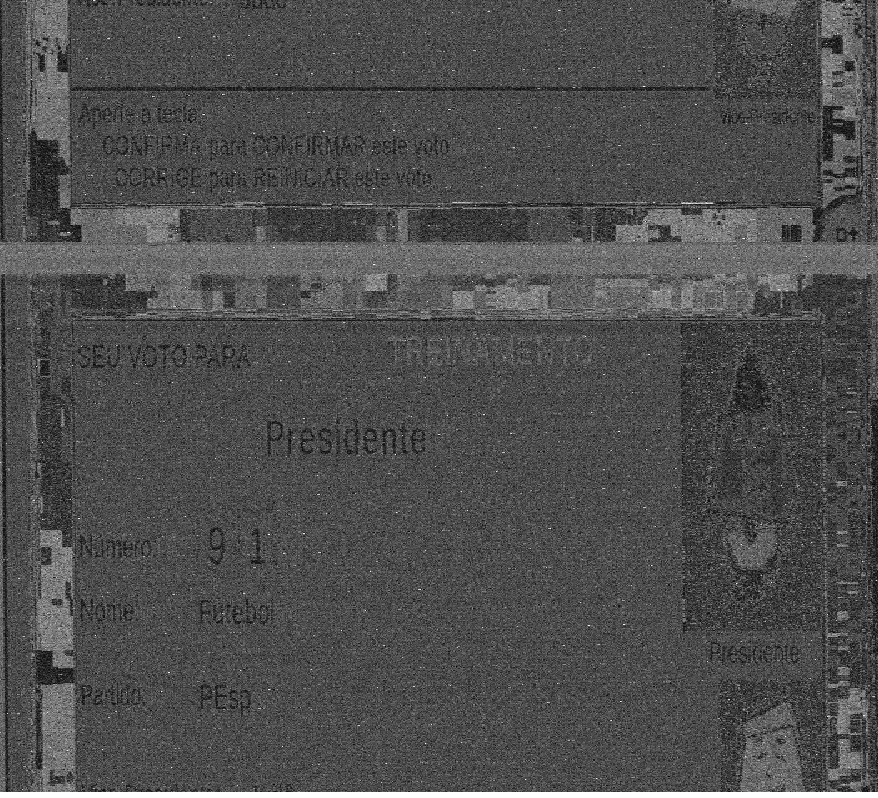}
 \end{figure}

Fig.~\ref{fig:91} shows an example of an e-vote frame reconstructed by \emph{Eve} 
from electromagnetic emanations captured from a VGA monitor operating at a resolution 
of $1280\times 720$ pixels and a refresh rate of 60~Hz. Prior knowledge of the screen 
resolution from public documents facilitates signal resynchronization. Once the remaining 
display timing parameters are determined, the reconstructed frames can be displayed as a 
real-time video, allowing \emph{Eve} to observe changes that \emph{Alice} makes to the 
displayed selection before confirming her vote. Similar reconstruction results were obtained 
for all tested display configurations.

To produce the image, we
implemented an experimental setup representative of the display component of a
Brazilian e-polling station\footnote{The TSE Public Security Test is typically held in the year preceding
an election; its most recent edition was held in 2025 for the 2026 General
Elections. The authors plan to apply to the next edition, expected in 2027,
to conduct experiments using an actual Brazilian EVM.}. Specifically, we used conventional VGA/HDMI
displays to reproduce the official Brazilian e-voting interface, which is
publicly available online for training purposes~\cite{tse_voting_simulator}.
We evaluated target-display resolutions of $1280\times720@60$ and
$1920\times1080@60$. The former corresponds to the minimum display resolution
specified for the UE2020 EVM model~\cite{ue2020hardware}. For these two
resolutions, the total timing dimensions $P_x\times P_y$, including blanking
intervals, are $1650\times750$ and $2200\times1125$, respectively~\cite{vesa_dmt_2013}.
 We were able to reconstruct the displayed e-vote under both configurations.

To acquire the leaked video signal, we employed an Ettus USRP B200 configured
with a sampling rate of $f_s=54$~MS/s and connected to a conventional digital
HDTV antenna. We found this sampling rate sufficient to reconstruct the
sensitive image (i.e., the selected vote), while providing a practical
trade-off between image quality and computational complexity. In our
experiments, the radio antenna and the target display were separated by
approximately 45~cm and a masonry wall, reproducing a scenario in which
\emph{Eve} operates from a room adjacent to the e-polling location. The
eavesdropped signal was processed using the \texttt{gr-tempest} module for GNU
Radio~\cite{grtempest-2022}, which enabled the reconstructed frames to be
displayed as a real-time video. Preliminary experiments indicated that the best 
reconstruction was obtained with the antenna oriented horizontally, perpendicular 
to the insertion axis of the VGA/HDMI cable connector. With the connector pointing 
upward, as in our testbed monitor, this configuration was achieved by rotating the 
antenna $90^\circ$ from its upright position about the pivot at its base.

For the frame shown in Fig.~\ref{fig:91}, \emph{Eve} sequentially inspected the first 
five harmonics of the pixel rate and identified a compromising component in the band 
centered at the fifth harmonic (371.25~MHz). Approximately 2~s were spent at each candidate frequency to assess 
whether a legible image could be obtained, resulting in a total search time of approximately
 10~s. Under the tested conditions, this initial search required only a small fraction of 
the time during which the polling station remains open on election day.

\section{Conclusion, Recommendations and Future Work}
This work investigated how a coordinated TEMPEST attack could compromise ballot secrecy by linking a reconstructed vote to a specific, non-colluding voter. The proposed scenario combines electromagnetic eavesdropping with observation of voter activity, extending the threat beyond vote reconstruction alone. Experiments using the official Brazilian e-voting interface displayed on a conventional monitor demonstrated through-wall vote reconstruction, the most critical stage of the proposed attack. These results support the feasibility of this stage under the tested conditions, but do not establish the vulnerability of official EVMs.

Our findings motivate consideration of procedural countermeasures that could hinder the association between voters and reconstructed votes. These include admitting voters into the polling room in groups, keeping the door closed between admissions except when needed for exit or assistance, and randomizing the voting order within each group. Such measures could reduce an external observer's ability to infer when a specific voter casts a ballot, although their effectiveness and operational feasibility require further evaluation.
In future work, we plan to investigate electromagnetic leakage from representative EVM hardware, evaluate the proposed coordinated attack under realistic polling conditions, and develop and assess signal-processing-based jamming countermeasures.

\bibliographystyle{IEEEtran}
\bibliography{IEEEabrv,refs}

\end{document}